\documentstyle[stwol]{article}

\def\be{\begin{equation}}
\def\ee{\end{equation}}
\def\bea{\begin{eqnarray}}
\def\eea{\end{eqnarray}}

\begin{document}

\thispagestyle{empty}

\title{
{\small\rm\vspace*{-2cm}
\rightline{TTP96-42}
\rightline{hep-ph/9609332}
\rightline{September 1996}
\ \\
\ \\}
CP VIOLATION AND CKM PHASES FROM TIME-DEPENDENCES OF UNTAGGED
$B_s$ DECAYS
\footnotemark[1]
}

\author{ROBERT FLEISCHER}

\address{Institut f\"ur Theoretische Teilchenphysik, Universit\"at 
Karlsruhe, D--76128 Karlsruhe, Germany}



\twocolumn[\maketitle\abstracts{
The $B_s$ system is analyzed in light of a possible width difference
$\Delta\Gamma_s$ between its mass eigenstates. If $\Delta\Gamma_s$ is 
sizable, {\it untagged} $B_s$-meson decays may allow a probe of CP 
violation and moreover the extraction both of the Wolfenstein parameter 
$\eta$ and of the notoriously difficult to measure angle $\gamma$ of 
the unitarity triangle. To accomplish this ambitious task, time-dependent 
angular distributions for untagged $B_s$ decays into admixtures of CP 
eigenstates and channels that are caused by $\bar b\to\bar c u\bar s$ 
quark-level transitions play a key role. The work described here was 
done in collaboration with Isard Dunietz.}]

\footnotetext[1]{Invited talk given at the {\it 
XXVIII International Conference on High Energy Physics -- ICHEP'96}, 
Warsaw, Poland, July 25--31, 1996, to appear in the proceedings.}

\section{Introduction}\label{intro}
The time-evolution due to $B^0_s-\overline{B^0_s}$ mixing is governed 
by the $B_s$ mass eigenstates $B_s^{\mbox{{\scriptsize Heavy}}}$ 
and $B_s^{\mbox{{\scriptsize Light}}}$ which are characterized by 
their mass eigenvalues $M_H^{(s)}$, $M_L^{(s)}$ and decay widths 
$\Gamma_H^{(s)}$, $\Gamma_L^{(s)}$. Because of these mixing effects, 
oscillatory $\Delta M_s t$ terms with $\Delta M_s\equiv M_H^{(s)}-
M_L^{(s)}$ show up in the time-dependent transition 
rates \cite{evol} $\Gamma(B^0_s(t)\to f)$ and $\Gamma(\overline{B^0_s}
(t)\to f)$ describing decays of initially present $B^0_s$ and 
$\overline{B^0_s}$ mesons into a final state $f$, respectively. 
The ``strength'' of the $B^0_s-\overline{B^0_s}$ oscillations is measured
by the mixing parameter $x_s\equiv\Delta M_s/\Gamma_s$, where
$\Gamma_s\equiv(\Gamma^{(s)}_H+\Gamma^{(s)}_L)/2$. Within the Standard 
Model one expects \cite{xs} $x_s={\cal O}(20)$ implying 
very rapid $B^0_s-\overline{B^0_s}$ oscillations 
which require an excellent vertex resolution system to keep track 
of the $\Delta M_s t$ terms. That is obviously a formidable experimental 
task.

However, as pointed out by Dunietz \cite{dunietz}, it may not be necessary 
to trace the rapid $\Delta M_s t$ oscillations in order to obtain insights 
into the mechanism of CP violation. This remarkable feature is due to the 
expected sizable width difference \cite{deltagamma} $\Delta\Gamma_s\equiv
\Gamma_H^{(s)}-\Gamma_L^{(s)}$. The major contributions to $\Delta\Gamma_s$, 
which may be as large as ${\cal O}(20\%)$ of the average decay width 
$\Gamma_s$, originate from $\bar b\to \bar cc\bar s$ transitions into final 
states that are common both to $B_s^0$ and $\overline{B_s^0}$. Because of
this width difference already {\it untagged} $B_s$ rates, which are defined 
by 
\begin{equation}\label{e1}
\Gamma[f(t)]\equiv\Gamma(B_s^0(t)\to f)+\Gamma(\overline{B^0_s}(t)\to f),
\end{equation}
may provide valuable information about the phase structure of the observable
\begin{equation}\label{e2}
\xi_f^{(s)}=\exp\left(-i\,\Theta_{M_{12}}^{(s)}\right)
\frac{A(\overline{B^0_s}\to f)}{A(B^0_s\to f)},
\end{equation}
where $\Theta_{M_{12}}^{(s)}$ is the weak $B_s^0$--$\overline{B_s^0}$ mixing
phase \cite{evol}. This can be seen nicely by writing Eq.~(\ref{e1}) in a 
more explicit way as follows:
\begin{eqnarray}
\lefteqn{\Gamma[f(t)]\propto\left[\left(1+\left|\xi_f^{(s)}
\right|^2\right)\left(e^{-\Gamma_L^{(s)} t}+e^{-\Gamma_H^{(s)} t}\right)
\right.}\nonumber\\
&&-2\mbox{\,Re\,}\xi_f^{(s)}\left(e^{-\Gamma_L^{(s)} t}-
e^{-\Gamma_H^{(s)} t}\right)\biggr].\label{e3}
\end{eqnarray}
In this expression the rapid oscillatory $\Delta M_s t$ terms, which
show up in the {\it tagged} rates, cancel~\cite{dunietz}. Therefore it 
depends only on the two exponents $e^{-\Gamma_L^{(s)} t}$ and 
$e^{-\Gamma_H^{(s)} t}$, where $\Gamma_L^{(s)}$ and $\Gamma_H^{(s)}$ can 
be determined e.g.\ from the angular distribution \cite{ddlr} of the decay
$B_s\to J/\psi\,\phi$. From an experimental point of view such {\it untagged}
analyses are clearly much more promising than tagged ones in respect of 
efficiency, acceptance and purity. 

\section{A Transparent Example}\label{example}
In order to illustrate these untagged rates in more detail, let me discuss 
an estimate of the angle $\gamma$ of the usual ``non-squashed'' unitarity 
triangle~\cite{ut} of the Cabibbo--Kobayashi--Maskawa matrix \cite{ckm}
(CKM matrix) using {\it untagged} $B_s\to K^+K^-$ 
and $B_s\to K^0\overline{K^0}$ decays. This approach has been proposed 
very recently by Dunietz and myself~\cite{fd1}. Using the $SU(2)$ isospin
symmetry of strong interactions to relate the QCD penguin contributions to
these decays (electroweak penguins are color-suppressed in these modes
and thus play a minor role), we obtain
\begin{eqnarray}
\lefteqn{\Gamma[K^+K^-(t)]\propto |P'|^2\Bigl[\bigl(1-2\,|r|\cos\rho\,
\cos\gamma}
\nonumber\\
&&+|r|^2\cos^2\gamma\bigr)e^{-\Gamma_L^{(s)} t}+|r|^2\sin^2\gamma\, 
e^{-\Gamma_H^{(s)} t}\Bigr]\label{e4}
\end{eqnarray}
and
\begin{equation}\label{e5}
\Gamma[K^0\overline{K^0}(t)]\propto |P'|^2\,e^{-\Gamma_L^{(s)} t},
\end{equation}
where 
\begin{equation}\label{e6}
r\equiv|r|e^{i\rho}=\frac{|T'|}{|P'|}e^{i(\delta_{T'}-\delta_{P'})}.
\end{equation}
Here $P'$ denotes \cite{ghlrsu3} the $\bar b\to\bar s$ QCD penguin amplitude, 
$T'$ is the color-allowed $\bar b\to\bar uu\bar s$ tree amplitude, and 
$\delta_{P'}$ and $\delta_{T'}$ are the corresponding CP-conserving strong 
phases. In order to determine $\gamma$ from the untagged rates 
Eqs.~(\ref{e4}) and (\ref{e5}) we need an additional input that  
is provided by the $SU(3)$ flavor symmetry of strong
interactions. If we neglect the color-suppressed current-current
contributions to $B^+\to\pi^+\pi^0$ we find \cite{ghlrsu3}
\begin{equation}\label{e7}
|T'|\approx\lambda\,\frac{f_K}{f_\pi}\,\sqrt{2}\,|A(B^+\to\pi^+\pi^0)|,
\end{equation}
where $\lambda$ is the Wolfenstein parameter \cite{wolf}, $f_K$ and
$f_\pi$ are the $K$ and $\pi$ meson decay constants, respectively,
and $A(B^+\to\pi^+\pi^0)$ denotes the appropriately normalized 
$B^+\to\pi^+\pi^0$ decay amplitude. Since $|P'|$ is known from
$B_s\to K^0\,\overline{K^0}$, the quantity $|r|=|T'|/|P'|$ can be estimated 
with the help of Eq.~(\ref{e7}) and allows the extraction of $\gamma$
from the part of Eq.~(\ref{e4}) evolving with the exponent 
$e^{-\Gamma_H^{(s)} t}$.

\section{$B_s$ Decays into Admixtures of CP Eigenstates}\label{admixtures}
As we will see in a moment, one can even do better than in the previous 
section, i.e.\ without using an $SU(3)$ flavor symmetry input, by 
considering the decays corresponding to $B_s\to K \overline{K}$ where 
two vector mesons (or higher resonances) are present in the final 
states \cite{fd1}.  

\subsection{An Extraction of $\gamma$ using Untagged $B_s\to K^{\ast+}
K^{\ast-}$ and $B_s\to K^{\ast0}\overline{K^{\ast0}}$ Decays}\label{kkbar}
The untagged angular distributions of these decays, which take the general 
form \cite{dighe}
\begin{equation}\label{e8}
[f(\theta,\phi,\psi;t)]=\sum_k\left[\overline{b^{(k)}}(t)+b^{(k)}(t)\right]
g^{(k)}(\theta,\phi,\psi),
\end{equation}
provide many more observables than the untagged modes
$B_s\to K^+K^-$ and $B_s\to K^0\overline{K^0}$ discussed in 
Section~\ref{example}. Here $\theta$, $\phi$ and $\psi$ are generic
decay angles describing the kinematics of the decay products arising in the
decay chain $B_s\to K^\ast(\to\pi K)\,\overline{K^\ast}(\to \pi\overline{K})$.
The observables $\left[\overline{b^{(k)}}(t)+b^{(k)}(t)\right]$ governing the
time-evolution of the angular distribution Eq.~(\ref{e8}) are given
by real or imaginary parts of bilinear combinations of decay amplitudes
that are of the following structure:
\begin{eqnarray}
\lefteqn{\left[A_{\tilde f}^\ast(t)\,A_f(t)\right]\equiv\left\langle
\left(K^\ast\overline{K^\ast}\right)_{\tilde f}\left|{\cal 
H}_{\mbox{{\scriptsize eff}}}\right|\overline{B_s}(t)\right\rangle^\ast}
\label{e9}\\
&&\times\left\langle\left(K^\ast\overline{K^\ast}\right)_{f}\left|{\cal 
H}_{\mbox{{\scriptsize eff}}}\right|\overline{B_s}(t)\right\rangle+
\left(\overline{B_s}\to B_s\right).\nonumber
\end{eqnarray}
In this expression $f$ and $\tilde f$ are labels that define the relative
polarizations of $K^\ast$ and $\overline{K^\ast}$ in final state 
configurations $\left(K^\ast\overline{K^\ast}\right)_f$ (e.g.\ linear 
polarization states \cite{rosner} $\{0,\parallel,\perp\}$) with CP 
eigenvalues $\eta_{\mbox{{\tiny CP}}}^f$:
\begin{equation}\label{e10}
({\cal CP})\left|\left(K^\ast\overline{K^\ast}\right)_f\right\rangle
=\eta_{\mbox{{\tiny CP}}}^f\left|
\left(K^\ast\overline{K^\ast}\right)_f\right\rangle.
\end{equation}
An analogous relation holds for $\tilde f$. The observables of the 
angular distributions for $B_s\to K^{\ast+}
K^{\ast-}$ and $B_s\to K^{\ast0}\overline{K^{\ast0}}$
are given explicitly in Ref.~\cite{fd1}. In the case of the latter decay
the formulae simplify considerably since it is a penguin-induced 
$\bar b\to\bar sd\bar d$ mode and receives therefore no tree contributions.
Using -- as in Section~\ref{example} -- the $SU(2)$ isospin symmetry of 
strong interactions, the QCD penguin contributions of these decays can be 
related to each other. If one takes into account these relations and goes
very carefully through the observables of the angular distributions, one 
finds that they allow the extraction of the CKM angle 
$\gamma$ {\it without} any additional theoretical input \cite{fd1}. 
In particluar no $SU(3)$ symmetry arguments 
as in Section~\ref{example} are needed.  The angular distributions provide 
moreover information about the hadronization dynamics of the corresponding
decays, and the formalism \cite{fd1} developed for $B_s\to K^{\ast+}
K^{\ast-}$ applies also to $B_s\to\rho^0\phi$ if we perform a suitable 
replacement of variables. Since that channel is expected to be dominated 
by electroweak penguins \cite{ewp}, it may allow interesing insights into 
the physics of these operators.

\subsection{The ``Gold-plated'' Transitions to Extract $\eta$}\label{gold}

This subsection is devoted to an analysis \cite{fd1} of the {\it untagged}
decays $B_s\to D_s^{\ast+}D_s^{\ast-}$ and $B_s\to J/\psi\,\phi$, which is 
the counterpart of the ``gold-plated'' mode $B_d\to J/\psi\,
K_{\mbox{{\tiny S}}}$ to measure the angle $\beta$ of the unitarity 
triangle. These decays are dominated by a single CKM amplitude. 
Consequently the hadronic uncertainties cancel in the quantity 
$\xi_f^{(s)}$ defined by Eq.~(\ref{e2}), which takes in that particular 
case the form
\begin{equation}\label{e11}
\xi_f^{(s)}=\exp(i\,\phi_{\mbox{{\tiny CKM}}}), 
\end{equation}
and the observables of the angular distributions simplify considerably. A
characteristic feature of these angular distributions is {\it 
interference} between
CP-even and CP-odd final state configurations leading to observables
that are proportional to  
\begin{equation}\label{e12}
\left(e^{-\Gamma_L^{(s)}t}-e^{-\Gamma_H^{(s)}t}
\right)\sin\phi_{\mbox{{\tiny CKM}}}.
\end{equation}
Here the CP-violating weak phase is given by \cite{xs} $\phi_{\mbox{{\tiny 
CKM}}}=2\lambda^2\eta\approx{\cal O}(0.03)$, where the Wolfenstein 
parameter $\eta$ fixes the height of the unitarity 
triangle \cite{ut}. The observables of the angular distributions \cite{fd1} 
for both the color-allowed channel $B_s\to D_s^{\ast+} D_s^{\ast-}$ and 
the color-suppressed transition $B_s\to J/\psi\,\phi$ each provide  
sufficient information to determine the CP-violating weak phase 
$\phi_{\mbox{{\tiny CKM}}}$ from their {\it untagged} data samples thereby 
fixing the Wolfenstein parameter $\eta$. The extraction of 
$\phi_{\mbox{{\tiny CKM}}}$ is not as clean as that of $\beta$ from 
$B_d\to J/\psi\,K_{\mbox{{\tiny S}}}$. This is due to the smallness of 
$\phi_{\mbox{{\tiny CKM}}}$ with respect to $\beta$ enhancing the 
importance of the unmixed amplitudes proportional to the CKM factor 
$V_{ub}^\ast V_{us}$ which are similarly suppressed in both cases. 

\section{$B_s$ Decays caused by $\bar b\to\bar cu\bar s$}\label{nonCP}
The $B_s$ decays discussed in this section are pure tree decays 
and probe the CKM angle $\gamma$ in a {\it clean} way~\cite{gam}. 
There are by now well-known strategies on the market using the 
time evolutions of such modes, e.g.\ $\stackrel{{\mbox{\tiny 
(---)}}}{B_s}\to\stackrel{{\mbox{\tiny 
(---)}}}{D^0}\phi$~\cite{gam,glgam} and $\stackrel{{\mbox{\tiny 
(---)}}}{B_s}\to D_s^\pm K^\mp$ \cite{adk}, to extract $\gamma$. However, in
these strategies {\it tagging} is essential and the rapid $\Delta M_s t$
oscillations have to be resolved which is an experimental challenge.
The question what can be learned from {\it untagged} data samples of 
these decays, where the $\Delta M_s t$ terms cancel, has been investigated 
by Dunietz in Ref.~\cite{dunietz}. In the untagged case the determination of 
$\gamma$ requires additional inputs: a measurement of the untagged $B_s\to 
D^0_{\mbox{{\tiny CP}}} \phi$ rate in the case of the color-suppressed 
modes $\stackrel{{\mbox{\tiny (---)}}}{B_s}\to\stackrel{{\mbox{\tiny 
(---)}}}{D^0}\phi$, and a theoretical input corresponding to the
ratio of the unmixed rates $\Gamma(B^0_s\to D_s^-K^+)/\Gamma(B^0_s\to
D_s^-\pi^+)$ in the case of the color-allowed decays
$\stackrel{{\mbox{\tiny (---)}}}{B_s}\to D_s^\pm K^\mp$. This ratio can 
be estimated with the help of the ``factorization'' hypothesis which may 
work reasonably well for these color-allowed channels.

Interestingly the {\it untagged} data samples may exhibit CP-violating 
effects that are described by observables of the form
\begin{equation}\label{e13}
\Gamma[f(t)]-\Gamma[\overline{f}(t)]\propto\left(e^{-\Gamma_L^{(s)}t}-
e^{-\Gamma_H^{(s)}t}\right)\sin\varrho\,\sin\gamma.
\end{equation}
Here $\varrho$ is a CP-conserving strong phase shift and $\gamma$ is the 
usual angle of the unitarity triangle. Because of the $\sin\varrho$ factor,
a non-trivial strong phase shift is essential in that case.
Consequently the CP-violating observables Eq.~(\ref{e13}) vanish within 
the factorization approximation predicting $\varrho\in\{0,\pi\}$. 
Since factorization may be a reasonable working assumption 
for the color-allowed modes $\stackrel{{\mbox{\tiny 
(---)}}}{B_s}\to D_s^\pm K^\mp$, the CP-violating effects in their
untagged data samples are expected to be very small. On the other hand,
the factorization hypothesis is very questionable for 
the color-suppressed decays $\stackrel{{\mbox{\tiny (---)}}}{B_s}\to
\stackrel{{\mbox{\tiny (---)}}}{D^0}\phi$ and sizable CP violation may 
show up in the corresponding untagged rates \cite{dunietz}. 

Concerning such CP-violating effects and the extraction of $\gamma$ from
{\it untagged} rates, the decays $\stackrel{{\mbox{\tiny (---)}}}{B_s}\to 
D_s^{\ast\pm} K^{\ast\mp}$ and $\stackrel{{\mbox{\tiny (---)}}}{B_s}\to
\stackrel{{\mbox{\tiny(---)}}}{D^{\ast0}}\phi$ are expected to be 
more promising than the transitions discussed above. As was
shown in Ref.~\cite{fd2}, the time-dependences of their untagged angular
distributions allow a {\it clean} extraction of the CKM angle $\gamma$
{\it without} any additional input. The final state configurations 
of these decays are not admixtures of CP eigenstates as in 
Section~\ref{admixtures}. They can, however, be classified by their parity
eigenvalues. A characteristic feature of the angular distributions are
interferences between parity-even and parity-odd configurations that may
lead to potentially large CP-violating effects in the untagged data samples
even when all strong phase shifts vanish. An example of such an
{\it untagged} CP-violating observable is the following quantity \cite{fd2}:
\begin{eqnarray}
\lefteqn{\mbox{Im}\left\{\left[A_f^\ast(t)\,A_\perp(t)\right]\right\}+
\mbox{Im}\left\{\left[A_f^{\mbox{{\tiny C}}\ast}(t)\,
A_\perp^{\mbox{{\tiny C}}}(t)\right]\right\}}\nonumber\\
&&\propto\left(e^{-\Gamma_L^{(s)}t}-e^{-\Gamma_H^{(s)}t}\right)
\bigl\{|R_f|\cos(\delta_f-\vartheta_\perp)\nonumber\\
&&+|R_\perp|\cos(\delta_\perp-\vartheta_f)\bigr\}\,\sin\gamma.\label{e14}
\end{eqnarray}
In that expression bilinear combinations of certain decay amplitudes 
(see Eq.~(\ref{e9})) show up, $f\in\{0,\parallel\}$ denotes a 
linear polarization state \cite{rosner} and $\delta_f$, $\vartheta_f$
are CP-conserving phase shifts that are induced through strong final 
state interaction effects. For the details concerning the 
observable Eq.~(\ref{e14}) -- in particular the definition of the 
relevant charge-conjugate amplitudes $A_f^{\mbox{{\tiny C}}}$ and the 
quantities $|R_f|$ -- the reader is referred to Ref.~\cite{fd2}. 
Here I would like to emphasize only that the strong phase
shifts enter in the form of {\it cosine} terms. Therefore non-trivial 
strong phases are -- in contrast to Eq.~(\ref{e13}) -- not essential for 
CP violation in the corresponding untagged data samples and one
expects, even within the factorization approximation, which may apply to
the color-allowed modes $\stackrel{{\mbox{\tiny (---)}}}{B_s}\to 
D_s^{\ast\pm} K^{\ast\mp}$, potentially large effects. 
 
Since the soft photons in the decays $D_s^\ast\to D_s\gamma$, $D^{\ast0}
\to D^0\gamma$ are difficult to detect, higher resonances exhibiting
significant all-charged final states, e.g.\ $D_{s1}(2536)^+\to
D^{\ast+}K^0$, $D_1(2420)^0\to D^{\ast+}\pi^-$ with $D^{\ast+}\to 
D^0\pi^+$, may be more promising for certain detector configurations. 
A similar comment applies also to the mode $B_s\to D_s^{\ast+}D_s^{\ast-}$ 
discussed in Subsection~\ref{gold}.

\section{Conclusions}
The oscillatory $\Delta M_st$ terms arising from $B_s^0-\overline{B_s^0}$
mixing, which may be too rapid to be resolved with present vertex
technology, cancel in {\it untagged} rates of $B_s$ decays that depend 
therefore only on the two exponents $e^{-\Gamma_L^{(s)}t}$ and 
$e^{-\Gamma_H^{(s)}t}$. If the width difference $\Delta\Gamma_s$ is
sizable -- as is expected from theoretical analyses -- {\it untagged}
$B_s$ decays may allow the determination both of the CKM angle $\gamma$
and of the Wolfenstein parameter $\eta$ and may furthermore provide 
valuable insights into the mechanism of CP violation and the hadronization 
dynamics of the corresponding decays. To this end certain angular 
distributions may play a key role. 

Compared to the tagged case, such untagged measurements are much more 
promising in view of efficiency, acceptance and purity. A lot of 
statistics is required, however, and the natural place for these 
experiments seems to be a hadron collider. Obviously the feasibility 
of untagged strategies to extract CKM phases depends 
crucially on a sizable width difference $\Delta\Gamma_s$. Even 
if it should turn out to be too small for such untagged analyses, 
once $\Delta\Gamma_s\not=0$ has been established experimentally, 
the formulae developed in Refs.~\cite{fd1,fd2} have also 
to be used to determine CKM phases correctly from tagged measurements. 
In this sense we cannot lose and an exciting future concerning $B_s$
decays may lie ahead of us!

\section*{Acknowledgment}
I would like to thank Isi Dunietz for the most pleasant and enjoyable 
collaboration on the \mbox{topics} presented in this talk.

\end{document}